\documentclass[hyper]{JHEP3}
\usepackage{amsmath,amsfonts,epsfig} 

\newcommand{\comment}[1]{}

\newcommand{\myref}[1]{(\ref{#1})}

\newcommand{\nn}{\nonumber}

\newcommand{\ZZ}{{\mathbb{Z}}}  % integers
\newcommand{\CC}{{\mathbb{C}}}  % complex numbers
\newcommand{\RR}{{\mathbb{R}}}

\newcommand{\unit}{1\hspace{-3pt}\mbox{l}}
\newcommand{\half}{\frac{1}{2}}

\newcommand{\del}{\partial}
\newcommand{\ket}[1]{|\,#1\,\rangle}

\newcommand{\kket}[1]{|\,#1\,\rangle\!\rangle}
\newcommand{\bbra}[1]{\langle\!\langle\,#1\,|}

\title{D-branes in Nonabelian Orbifolds with Discrete Torsion} 

\author{Tam\'{a}s Hauer and Morten Krogh \\
  \\ CERN Theory Division 
  \\ CH -- 1211 Geneva 23 
  \\ Switzerland \\
  \\ E-mail: \email{tamas.hauer@cern.ch, morten.krogh@cern.ch}} 

\abstract{We study IIB string theory on the orbifold $\RR^8/\Gamma$
  with discrete torsion where $\Gamma$ is an arbitrary subgroup of
  $U(4)$.  We extend some previously known identities for discrete
  torsion in abelian groups to nonabelian groups.  We construct
  explicit formulas for a large class of fractional D-brane states and
  prove that the physical states are classified by projective
  representations of the orbifold group as predicted originally by
  Douglas.  The boundary states are found to be linear combinations of
  Ishibashi states with the coefficients being characters of the
  projective representations.}


\preprint{CERN-TH/2001-245 \\ hep-th/0109170}

\begin{document} 

% %%%%%%%%%%%%%%%%%%%%%%%%%%%%%%%%%%%%
%\setlength{\parskip}{2ex}
%\setlength{\parindent}{0em}
%\setlength{\baselineskip}{3ex}
% \pagestyle{myheadings}
%\markright{D-Branes and Projective Representations: \today}
%%%%%%%%%%%%%%%%%%%%%%%%%%%%%%%%%%%%

\section{Introduction}
\label{Introduction}

In this paper we will study certain orbifolds of type IIB string
theory with discrete torsion and characterize D-brane states in these
theories.  Type IIB in ten dimensions is already an orbifold, namely
of the GSO group. We further mod out by a geometrical orbifold group,
$\Gamma_0 \subset U(4) \subset SO(8)$ to get type IIB on
$\RR^8/\Gamma_0$.  For a given orbifold group there are various
possibilities for a consistent theory, these possibilities differ by
discrete torsion \cite{Vafa:1986wx,Vafa:1995rv}.  An effective
approach to studying D-branes in string theory -- which we will follow
-- is provided by the boundary state formalism
\cite{Callan:1988wz,Polchinski:1988tu}, a review of which can be found
e.g. in \cite{DiVecchia:1999rh,DiVecchia:1999fx,Recknagel:1998sb}.
The study of D-brane states in abelian orbifolds with discrete torsion
was initiated by Douglas and Fiol \cite{Douglas:1998xa,Douglas:1999hq}
who conjectured that D-branes were classified by projective
representations of the orbifold group.  Later more work was done in
various examples
\cite{Mukhopadhyay:1999pu,Berenstein:2000hy,Gomis:2000ej,
  Klein:2000tf,Aspinwall:2000xs,Aspinwall:2000xv,Feng:2000mw,
  Sharpe:2000ki} and this picture was confirmed in the abelian case.
For results on orbifolds without discrete torsion, see for example
\cite{Gutperle:2000bf,Brunner:1999fj,Diaconescu:1999dt,
  Diaconescu:2000ec,Billo:2000yb,Gaberdiel:1999ch}.

The aim of this paper is to describe the D-branes in nonabelian
orbifold theories with discrete torsion and prove that they are given
in terms of projective representations of the orbifold group.  In
order to do this we present explicit formulas for the relation between
discrete torsion in Vafa's sense and the cocycles of projective
representations.  
In particular we propose the following relation between the discrete
torsion and the cocycles in the general (nonabelian) case:
\begin{align*}
  \varepsilon(h,g) = \frac{c(ghg^{-1},g)}{c(g,h)}\,.
\end{align*}
We write down explicit formulas for the boundary states of the
D-branes, impose orbifold invariance and use open--closed string
duality to impose a physical condition.  We then show that this
condition is solved by characters of projective representations
exactly as predicted by Douglas.  It turns out that for $\Gamma_0
\subset U(4) \subset SO(8)$ certain boundary states can be written in
a uniform way that does not require fermionic zero modes to be treated
separately.  In order to classify the boundary states it is necessary
to fix various phases in the orbifold action on the closed string
Hilbert space.  We make the assumption, which was also employed in
\cite{Gaberdiel:2000fe,Craps:2001xw}, that one of the orbifold
theories contain all the fractional D-instantons and anti D-instantons
as in \cite{Douglas:1996sw}; in other words there should exist a
theory in which D-instantons are classified by ordinary
representations. This assumption partly determines the unknown phases
while the rest can be fixed by introducing discrete torsion in the
product of the GSO group and the geometrical group.  Having fixed
these phases the spectrum of open string states can be read off and we
prove that the consistency conditions are solved by projective
characters.  Our result is the following.  Let $\kket{h;S,\eta}$
denote the Ishibashi state in the $h$-twisted sector solving the
gluing condition $(S,\eta)$ (see the text for the precise definitions)
and let $\Gamma_S\subset\Gamma$ be the subgroup of the orbifold group
containing elements which leave the gluing condition $(S,\eta)$
invariant.  Then the following boundary states define D-branes:
\begin{align*}
  \nn \kket{R\,;S} &= \frac{1}{\sqrt{|\Gamma|}}
  \sum_{[f]\in\Gamma/\Gamma_{S}} \sum_{h\in\Gamma_{S}}
  \mbox{Tr}(\gamma_R(h))\,\varepsilon(h,f)\,f\kket{h;S,+} \,,
\end{align*}
where $R:h\mapsto\gamma_R(h)$ is a projective representation of
$\Gamma_S$.

The paper is organized as follows: 
In section \ref{Orbifoldthy} we specify the orbifold theories that
             we are interested in and discuss discrete torsion.
In section \ref{Boundarycovering} we construct boundary states
             explicitly and impose invariance of these under the
             orbifold group, both with and without discrete torsion.
In section \ref{Openclosed} we impose open string--closed string
             duality on the boundary states in order to determine the
             spectrum of physical states. We show that the spectrum is
             determined by projective representations of certain
             subgroups of the orbifold group.
In section \ref{Conclusion} we conclude.

\section{Orbifold theories}
\label{Orbifoldthy}

We shall determine boundary states in various orbifolds of superstring
theory. For simplicity we will specialize to type IIB theory. We start
with the ten-dimensional GSO-unprojected superstring and include the
GSO in the orbifold group.

We shall be working in light cone gauge with $0,1$ being the
light-cone coordinates and let the orbifold group act on the remaining
8 coordinates only. We require that the action of the elements of the
orbifold group $g\in\Gamma$ has the following form:
\begin{equation}  
  \label{fermionorbifold}
  \begin{split}
    X^\mu \rightarrow A^{\mu\nu}(g)X^\nu \qquad\qquad\qquad\qquad&
    \psi_{L}^\mu \rightarrow  -A_L^{\mu\nu}(g)\psi_{L}^\nu
    \\&
    \psi_{R}^\mu \rightarrow  -A_R^{\mu\nu}(g)\psi_{R}^\nu \,.
  \end{split}
\end{equation}
where $\mu=2\ldots 9$, $A(g)$ is an $U(4)\subset SO(8)$ matrix and
$A_L(g) , A_R(g)$ are equal to $A(g)$ up to a possible sign.  A
generic geometric orbifold could be obtained by choosing a subgroup
$\Gamma\subset SO(8)$.  In this paper however, we chose to restrict to
groups $\Gamma$ which respect a complex structure on $\RR^8$ thus are
embedded in a subgroup $U(4)\subset SO(8)$.  These orbifolds preserve
some spacetime supersymmetry and the use of the complex structure
makes our analysis more straightforward.  Also, we shall focus on
boundary states which respect this structure.  The transformation of
the fermions is determined by the transformation of the bosons from
worldsheet supersymmetry up to a sign.  For example the usual
geometric orbifold would correspond to $A_L =A_R = A$ while the four
elements used in the GSO projection have $A=\unit$.  We have also
included a minus sign for the fermions so that the NSNS sector
corresponds to the untwisted sector thus reflecting the periodicity on
the complex plane.

We will always impose the GSO projection, therefore the most general
form of the orbifold group is
\begin{align*}
  \Gamma &= \Gamma_0 \times 
  (\ZZ_{2}^{(L)}\times\ZZ_{2}^{(R)})_{GSO},
\end{align*}
where $\Gamma_0$ contains elements with $A_L =A_R = A$ in
\myref{fermionorbifold}. It is important to realize that the orbifold
is not completely defined by \myref{fermionorbifold} for the following
reason.  The orbifold is defined by first introducing twisted sectors,
one for each group element.  The quantization of these sectors gives
rise to a covering Hilbert space whose $\Gamma$ invariant subspace
defines the physical Hilbert space. To keep the $\Gamma$ invariant
subspace it is necessary to define an action of $\Gamma$ on each
twisted sector.  Each twisted sector is a Fock space where the tower
of states can be obtained by acting with the raising operators from
$X^{\mu}$ and $\psi^{\mu}$ on a lowest weight state.
\myref{fermionorbifold} defines the action of $\Gamma$ on the raising
operators but not on the lowest weight state.  In the case of zero
modes there are many possible choices of the lowest weight state, but
any choice will do, because the zero modes will transform these among
themselves.  To define the action of $\Gamma$ on the large Hilbert
space it is thus necessary to define the action of $\Gamma$ on each of
the lowest weight states. Furthermore this action has to be defined
such that $\Gamma$ is a symmetry of the theory, i.e.  $\Gamma$ is a
symmetry of the OPE. The $\Gamma$ invariant subspace then defines a
unitary, modular invariant string theory.

Two questions are inevitable at this point: does the orbifold exist at
all and is it unique? To our knowledge no one has proven that the
orbifold exists for all $\Gamma$, but in many special cases it is
known, of course. It could be checked explicitly but since it is out
of the main course of this paper we will refrain from this and just
assume that the orbifold exists.  Alternatively one could say that our
results about boundary states are only valid when the orbifold exists.

The uniqueness issue has been studied initially by Vafa
\cite{Vafa:1986wx}. In general it is not unique and the non-uniqueness
is classified by discrete torsion as we will describe in the next
subsection.

\subsection{Discrete Torsion}
\label{OrbifoldDT}

Given an orbifold of type IIB/$\Gamma$ one can possibly define other
related theories. The original orbifold is defined by having an action
of $\Gamma$ on the covering Hilbert space including all twisted
sectors and restricting to $\Gamma$-invariant states. The projection
operator onto the physical subspace is given by
\begin{align*}
  P = \frac{1}{|\Gamma|}\sum_{g \in \Gamma} g \,.
\end{align*} 
One can modify this theory by introducing discrete torsion
\cite{Vafa:1986wx} as follows. One defines new projection operators
for each twisted sector:
\begin{align*}
  P_h = \sum_{g \in \Gamma} \varepsilon(h,g) g \,,
\end{align*} 
where $h \in \Gamma$ is the twist and $\varepsilon(h,g)$ are complex
numbers of unit norm. The physical states are the ones that are
invariant under these projection operators.  Remark that in general
$P_h$ maps a state twisted by $h$ into a linear combination of states
twisted by elements in the same conjugacy class as $h$.

Another way of understanding discrete torsion is as a redefinition of
the elements of $\Gamma$. In the sector twisted by $h$ the
redefinition is given by $ \widehat{g} = \varepsilon(h,g)g$ with some
conditions on $\varepsilon$.  One comes from the fact that the
redefined group elements form a representation of $\Gamma$, i.e.
\begin{align*}
  \widehat{g_2 g_1} = \widehat{g_2} \widehat{g_1}
\end{align*}
Applying both sides to a state $\ket{h}$ which is twisted by $h$ and
remembering that $g_1 \ket{h}$ is twisted by $g_1 h g_1^{-1}$ one gets
that
\begin{align}
  \varepsilon(h,g_2g_1) = 
  \varepsilon(g_1hg_1^{-1},g_2)\varepsilon(h,g_1)\,,
  \label{disrep}
\end{align}
Another condition comes from modular invariance of the torus amplitude
\cite{Vafa:1986wx} which requires that $\varepsilon$ obey
\begin{align}
  \varepsilon(g,h) &= \varepsilon(g^a h^b , g^c h^d) ,
  \qquad\mbox{where}\qquad  
  \left(\begin{array}{cc} a & b \\ c & d\end{array}\right) \in
  SL(2,\ZZ) \qquad \mbox{and} \qquad gh=hg  \,.
  \label{modinv}
\end{align}
Note that this only applies to commuting elements $g,h$, because there
is no torus amplitude with noncommuting twist along the two
directions.  For noncommuting elements $g,h$, we have
\begin{align*}
  \widehat{g} \ket{h} = \varepsilon(h,g)g\ket{h} 
  = \varepsilon(h,g)\ket{ghg^{-1}} \,,
\end{align*}
where $\ket{k}$ is some state in the sector twisted by $k$. We see
that multiplying $\ket{h}$ by a phase changes $\varepsilon(h,g)$ by a
phase; $\varepsilon$ is thus basis dependent to a certain degree. Of
course $\varepsilon$'s that only differ in this way should be
identified.  For commuting elements this does not happen since both
sides of the equation contain $\ket{h}$ thus for abelian groups
$\varepsilon$ is independent of the choice of basis.

The set of $\widehat{g}$ constitute a group isomorphic to $\Gamma$
with another action on the Hilbert space but the same action on the
fields $X^{\mu}$ and $\psi^{\mu}$. It is thus not well defined to ask
whether an orbifold has discrete torsion. The correct statement is
that there are possibly several possibilities for the orbifold and
given one of them the others are obtained by introducing discrete
torsion.  Still we will talk about the orbifold without discrete
torsion whereby we mean one of these theories.

There is a relation between discrete torsion of $\Gamma$ and the
projective representations of $\Gamma$. A projective finite
dimensional representation of $\Gamma$ is a map, $\gamma$, from
$\Gamma$ into $GL(n,\CC)$ for some $n$ which obeys
\begin{align}
  \gamma(g) \gamma(h) = c(g,h) \gamma(gh) \,,
  \label{projdef}
\end{align}
where $c(g,h)$ are complex numbers of unit norm.  Because of the
associative law $c(g,h)$ satisfies
\begin{align} 
  c(g_1 , g_2 g_3) c(g_2 ,g_3) = c(g_1 ,g_2) c(g_1 g_2, g_3) \,.
  \label{projassoc}
\end{align}
By redefining $\gamma(g)$ with a phase $c_g$, we can change the
coefficients as
\begin{align}
  c(h,g) \rightarrow  \frac{c_h c_g}{c_{hg}} c(g,h) \,.
 \label{redef}
\end{align}
These relations define the cohomology group $H^2(\Gamma , U(1))$ whose
elements thus correspond to types of projective representations. For
each type there are many representations, for instance $1 \in
H^2(\Gamma , U(1))$ correspond to all the ordinary representations.

The relation between projective representations and $\varepsilon$ is
well known for abelian orbifolds: given an element in $H^2(\Gamma ,
U(1)))$ with a representative $c(g,h)$ one defines
\begin{align}
  \varepsilon(h,g) = \frac{c(h,g)}{c(g,h)}\,.  
  \label{cepsilonabel}
\end{align}
For nonabelian orbifolds however, this definition does not lead to
$\varepsilon$ which satisfies \myref{disrep}.  Instead we propose the
following generalization:
\begin{align}
  \varepsilon(h,g) = \frac{c(ghg^{-1},g)}{c(g,h)}\,,
  \label{cepsilon}
\end{align}
which reduces to \myref{cepsilonabel} in the case of commuting
elements.  We now use \myref{projassoc} to show that $\varepsilon$ so
defined satisfies the condition \myref{disrep}:
\begin{align*}
  \varepsilon(h,g_2g_1) 
  &= \frac{c(g_2g_1hg_1^{-1}g_2^{-1}, g_2g_1)}{c(g_2g_1,h)}  
  = \frac{c(g_2g_1hg_1^{-1}, g_1) 
    c(g_2g_1hg_1^{-1}g_2^{-1},g_2)}{c(g_2,g_1)c(g_2g_1,h)} 
  \\ &= \frac{c(g_2, g_1h) c(g_1hg_1^{-1},g_1) 
    c(g_2g_1hg_1^{-1}g_2^{-1},g_2)}
  {c(g_1,h)c(g_2,g_1h)c(g_2,g_1hg_1^{-1})} 
  = \varepsilon(g_1hg_1^{-1}, g_2) \varepsilon(h,g_1) \,.
\end{align*}
We also see from \myref{cepsilon} that $\varepsilon(h,h)=1$ and for
commuting elements
\begin{align*}
  \varepsilon(h,g)=\varepsilon(g,h)^{-1}= \varepsilon(g,h^{-1}) 
  \qquad\qquad\qquad gh=hg\,.
\end{align*}
All this implies the modular invariance condition \myref{modinv}.

It is important to note that $\varepsilon$ depends on the
representative $c(g,h)$ and not just on the cohomology class.  Under
the redefinition \myref{redef} $\varepsilon(h,g)$ changes as
\begin{align}
  \varepsilon(h,g) \rightarrow  \frac{c_{ghg^{-1}}}{c_{h}} 
  \varepsilon(h,g)\,,
  \label{epsredef}
\end{align}
which is in good accord with the group action enhanced by discrete
torsion: 
\begin{align*}
  \ket{ghg^{-1}} = \varepsilon(h,g) g \ket{h} \,,
\end{align*}
where $\ket{k}$ is a state in the sector twisted by $k$.  Redefining
states in the $k$-twisted sector by $c_k$ the above equation will lead
to a redefinition of $\varepsilon$ exactly as in \myref{epsredef}.
This is a novel feature in the nonabelian case: when $hg=gh$ the left
and right hand sides change in the same way and $\varepsilon$ is
invariant.  Good discussions of discrete torsion in the abelian case
can be found among others in \cite{Gaberdiel:2000fe}.

\subsection{Consistency and fractional D-branes}
\label{ConsistencyDb}

We are interested in understanding the structure of the D-brane states
in all these orbifold theories.  In order to do this it is necessary
to know the exact action of $\Gamma$ on the states in the covering
Hilbert space.  This is related to the question about the existence of
the orbifold. A complete analysis of the existence would fix the
complete action of $\Gamma$.  We will however make one further
assumption -- first used by Bergman and Gaberdiel
\cite{Bergman:2000ni} in some specific cases -- which fixes the
action.  We assume that for each group $\Gamma$, there is an orbifold,
type IIB / $\Gamma$, in which all fractional D-instantons exist at the
orbifold fixed point. By all fractional branes we mean a brane and an
antibrane corresponding to each irreducible representation of
$\Gamma_0$.  Below it will become clear that it partly fixes the
action of $\Gamma$.

In summary we are making the following two assumptions that could, 
in principle, be checked:
\begin{enumerate}
\item  type IIB / $\Gamma$ exists
\item  This orbifold, or at least one of them in 
  the case of non trivial discrete 
  torsion, has a fractional D-instanton (and anti D-instanton) 
  for each representation of $\Gamma_0$. We will a bit imprecisely 
  say that this orbifold is without discrete torsion.    
\end{enumerate}

\section{Boundary states on the covering space} 
\label{Boundarycovering}

We are interested in classifying the structure of boundary states
describing D-branes in the orbifolded theory. We will consider
D-branes which touch the fixed point of the orbifold point group
($X^{\mu} = 0$). D-branes that do not touch $X^{\mu} = 0$ do not have
as interesting a structure; they are just the same D-branes as in the
parent theory with their images.  For definiteness we also take the
light-cone directions to be Dirichlet.

The closed string Hilbert space is a sum of twisted sectors which are
labeled by conjugacy classes of $\Gamma$ and only $\Gamma$-invariant
states are retained. In fact it is $\Gamma$-invariance that collects
twists into conjugacy classes.  A boundary state is a closed string
state which obeys certain gluing conditions corresponding to the
boundary conditions in the dual open string picture. A convenient way
to construct these states is to work on the covering space (the
unprojected Hilbert space) and impose $\Gamma$-invariance afterwards.

The orbifold group is a symmetry of the unprojected Hilbert space
which is thus represented by unitary operators there:
\begin{align}
  g^{-1} X^{\mu}(\tau,\sigma) g &= A(g)^{\mu\nu} X^{\nu}(\tau,\sigma)
  &
  g^{-1} \psi_L^{\mu}(\tau,\sigma) g &= 
  A_L(g)^{\mu\nu} \psi_L^{\nu}(\tau,\sigma)
  \nn \\ && 
  \label{orbihilbert}
  g^{-1} \psi_R^{\mu}(\tau,\sigma) g &= 
  A_R(g)^{\mu\nu} \psi_R^{\nu}(\tau,\sigma)
  \\
  A(gh) &= A(g) A(h) \,.
  \nn
\end{align}
In the sector twisted by $h$ we require 
\begin{equation}
  \label{twist}
  \begin{split}
    X^\mu(\sigma + 2\pi) = A(h)^{\mu \nu} X^{\nu}(\sigma) 
    \qquad\qquad\qquad\qquad 
    \psi^\mu_L(\sigma + 2\pi) &= 
    - A_L^{\mu\nu}(h) \psi^\nu_L(\sigma) 
    \\ 
    \psi^\mu_R(\sigma + 2\pi) &= 
    - A_R^{\mu\nu}(h) \psi^\nu_R(\sigma) \,.
  \end{split}
\end{equation}
Combining \myref{orbihilbert} and \myref{twist} shows that $g$ maps
from the $h$-twisted sector into the $ghg^{-1}$-twisted sector:
\begin{align*}
%  (X^{\mu}(\tau,\sigma+2\pi)- 
%  A(h)^{\mu\nu} X^{\nu}(\tau,\sigma))\ket{h} &= 0
%  \\
  (gX^{\mu}(\tau,\sigma+2\pi)g^{-1}- 
  gA(h)^{\mu\nu} X^{\nu}(\tau,\sigma)g^{-1})g\ket{h} &= 0
  \\
%  (A^{\mu\nu}(g^{-1})X^{\nu}(\tau,\sigma+2\pi)- 
%  A(h)^{\mu\rho}A^{\rho\nu}(g^{-1}) X^{\nu}(\tau,\sigma))g\ket{h} &= 0
%  \\
  (X^{\mu}(\tau,\sigma+2\pi)- 
  A(ghg^{-1})^{\mu\nu} X^{\nu}(\tau,\sigma))g\ket{h} &= 0 \,.
\end{align*}

Let us now describe the boundary states on the covering space, i.e.
before projecting onto $\Gamma$-invariant states.  They are
characterized by the gluing condition at the edge of the string
worldsheet.  As we mentioned in the introduction, we chose to study
those boundary state which respect a complex structure which is left
untouched by the orbifold.  Thus the gluing condition is given in
terms of a matrix $S\in U(4)$ and a number $\eta \in \{-1,1\}$ where
$U(4)\subset SO(8)$ is the same subgroup as the one appearing in the
definition of the orbifold:
\begin{align}
  \begin{cases}
    \qquad(\del_L X^\mu(\tau=0,\sigma) + 
    S^{\mu\nu}\del_R X^\nu(\tau=0,\sigma))  
    \kket{h,S,\eta} &= 0 
    \\
    \qquad(\psi^\mu_L(\tau=0,\sigma) 
    + i \eta S^{\mu\nu}\psi^\nu_R(\tau=0,\sigma))   
    \kket{h,S,\eta} &= 0. 
  \end{cases}
  \label{ishibashiclosed}
\end{align}
We denote the boundary state as $\kket{h,S,\eta}$, where $h \in
\Gamma$ is the twist of the state and $S , \eta$ are the parameters in
the gluing condition.  In order to conform with the local worldsheet
supersymmetry the gluing condition of the fermions is determined by
that of the bosons up to the sign $\eta$.

It turns out that for a given gluing condition a boundary state only
exists in certain twisted sectors. Let us show how this comes about.
Plugging \myref{twist} into \myref{ishibashiclosed} and rearranging we
get
\begin{align*}
  (\del_L X^\mu(\tau=0,\sigma) + 
  (A^{-1}(h)SA(h))^{\mu\nu}\del_R X^\nu(\tau=0,\sigma))  
  \kket{h;S,\eta} &= 0 
  \\
  (\psi_L^\mu(\tau=0,\sigma) + i\eta
  (A_L^{-1}(h)SA_R(h))^{\mu\nu}\psi_R^\nu(\tau=0,\sigma))  
  \kket{h;S,\eta} &= 0 \,.
\end{align*}
It is easy to show that the only simultaneous solution to these
equations and \myref{ishibashiclosed} is zero unless
\begin{align*}
  S &= A^{-1}(h)SA(h) 
  &
  S &= A_L^{-1}(h)SA_R(h) \,,
\end{align*}
showing that $A(h)$ must commute with $S$.  As $A_L(h),A_R(h)$ are
equal to $A(h)$ up to a sign we conclude that the twist must be in the
following subgroup of $\Gamma$:
\begin{align*}
  \Gamma_{sym} &= \{A_L(h)= A_R(h)=A(h)\}\cup
  \{A_L(h)= A_R(h) = - A(h)\}\,.  
\end{align*} 
This means in particular that the boundary state only exists in the
NSNS- or RR-sector but not in the RNS- and NSR-sectors.

Now we will present explicit solutions to the gluing conditions in
terms of creation operators which apper in the oscillator expansion of
the fields.  The boundary state is a product of a solution to the
bosonic and fermionic equations. First we will consider the bosonic
part.  The coordinates in the sector twisted by $h$ satisfy
\myref{twist} where $A^{\mu \nu} = A^{\mu\nu}(h)$ is a $8 \times 8$
matrix in $U(4)\subset SO(8)$. We suppress the $h$ dependence of
$A^{\mu \nu}$ for the moment.  $A^{\mu \nu}$ has 8 complex eigenvalues
of unit norm, coming in pairs, which we parameterize by four phases
$\alpha_i\in[0,2\pi)$ :
\begin{align}
  A^{\mu\nu}v_i^\nu &= e^{i\alpha_i}v_i^\mu 
  &
  A^{\mu\nu}{v_i^\nu}^* &= e^{-i\alpha_i}{v_i^\mu}^*
  &
  \sum_\mu v_i^\mu {v_j^\mu}^* &= \delta_{ij}.
  & i&=1\ldots 4 \,.
  \label{orthogonal}
\end{align} 
The equation $X^\mu(\sigma +2\pi) = A^{\mu\nu} X^\nu(\sigma)$
possesses the following complete set of solutions:
\begin{align} 
  f_{i,n}^\mu(\sigma) &= e^{i(\frac{\alpha_i}{2\pi}+n)\sigma}v_i^\mu
  &
  f_{i,n}^{\mu*}(\sigma) &=
  e^{-i(\frac{\alpha_i}{2\pi}+n)\sigma}v_i^{\mu*} 
  & n\in \ZZ\,;\;\; i=1\ldots 4 \,.
  \label{ffunctions}
\end{align} 
We need to treat zero modes a bit different. Let $I_0 \subset
\{1,2,3,4\}$ such that $\alpha_i = 0 $ if and only if $i \in I_0$.
Then $X^\mu$ can be expressed as a mode expansion :
\begin{align*}
  X^\mu(\sigma, \tau) &= 
  \sum_{i \in I} (p^i \tau + x^i) v_i^{\mu} + \\ &
  \sum_{\substack{ n \in \ZZ ,i \in \{1,2,3,4\} \\
    \mbox{\footnotesize{or }} n=0 , i \not\in  I_0}}
  \frac{\alpha_n^if^\mu_{i,n}(\sigma-\tau)}
       {i (\frac{\alpha_i}{2 \pi}+n)} +
  \frac{ \alpha_n^{i\dagger} f_{i,n}^{\mu*}(\sigma-\tau)}
       {-i (\frac{\alpha_i}{2 \pi} +n)}
  +
  \frac{\tilde\alpha_{-n}^i f^\mu_{i,n}(\sigma+\tau)}
       {i (\frac{\alpha_i}{2 \pi} +n)} +
  \frac{\tilde\alpha_{-n}^{i\dagger} f_{i,n}^{\mu*}(\sigma+\tau)}
  {-i (\frac{\alpha_i}{2 \pi} +n)} \,.
\end{align*}
The commutation relations are as follows:
\begin{align*}
  [ x^i , p^j ] &= i \delta^{ij}
  &
  [\alpha^i_n , \alpha^{j \dagger}_n ] &= 
  \half (\frac{\alpha_i}{2 \pi} +n) \delta^{ij}
  &
  [\tilde \alpha^i_{-n} , \tilde \alpha^{j \dagger}_{-n} ] &= 
  - \half (\frac{\alpha_i}{2 \pi} +n)\delta^{ij} \,.
\end{align*}
All other commutators are zero. $\alpha^i_{n\geq 0},
\tilde\alpha^i_{n>0}$ are annihilation operators and $\alpha^i_{n< 0},
\tilde\alpha^i_{n\leq 0}$ are creation operators, for $n=0$ these
operators exist for $i \not\in I_0$ only.

The gluing condition for the bosonic part reads 
\begin{align*}  
  (\del_L X^\mu(\tau=0,\sigma) + 
  S^{\mu\nu}\del_R X^\nu(\tau=0,\sigma)) \kket{h;S,\eta} &= 0 \,.
\end{align*}
Plugging the mode expansion of $X$ into the gluing condition and using
the linear independence of $f_{i,n}, f^*_{i,n}$ one arrives at the
following set of equations:
\begin{align*}
  \sum_{i \in I_0} \left(  p^i v_i^{\mu} - p^i S^{\mu \nu} v_i^{\nu} 
  \right)  \kket{h;S,\eta} &= 0 
  \\
  \left(
    \tilde\alpha_{-n}^if_{i,n}^\mu(\sigma)+
    \alpha_n^iS^{\mu\nu}f^\nu_{i,n}(\sigma)
  \right)\kket{h;S, \eta} &= 0
  && \forall\; i,n 
  \\ 
  \left(
    \tilde\alpha_{-n}^{i\dagger} f_{i,n}^\mu(\sigma)^* +
    \alpha_n^{i\dagger}S^{\mu\nu}f^\nu_{i,n}(\sigma)^*
  \right)\kket{h;S, \eta} &= 0
  && \forall\; i,n  \,.
\end{align*}
Here we also used that $S$ respects the complex structure used to
define the eigenvectors in \myref{orthogonal}. In other words $S$ maps 
$v_i$ into a linear combination of $v_j$ and not into $v_j^*$.    
The first equation
essentially fixes the overall momentum of the boundary state. In the
generic case, where $S^{\mu \nu}$ does not have 1 as eigenvalue, it
implies
\begin{align*}
  p^i \kket{h;S, \eta} = 0 \,.
\end{align*}
In other cases other momenta are possible corresponding to the fact
that the D-brane can move in certain directions. At the moment we will
solve the zero mode part of the gluing condition by putting $p^i =0$.
The ground state of the Fock space in the sector twisted by $h$ with
zero momentum is denoted by $\ket{h}$. Of course, in some twisted
sectors there are no momenta, in this case $\ket{h}$ just denotes the
ground state.

The equations for the non-zero modes can be rewritten as
\begin{align*}
  \left(
    \tilde\alpha_{-n}^i v_{i}^\mu+
    \alpha_n^iS^{\mu\nu} v^\nu_{i}
  \right)\kket{h;S, \eta} &= 0
  && \forall\; i,n 
  \\ 
  \left(
  \tilde\alpha_{-n}^{i\dagger} v_{i}^{\mu*} +
  \alpha_n^{i\dagger}S^{\mu\nu} v^{\nu*}_{i}
  \right)\kket{h;S, \eta} &= 0
  && \forall \; i,n  \,.
\end{align*}
Using the orthogonality \myref{orthogonal} of $v_i^\mu$ we get
\begin{equation}
  \begin{split}
  ( \tilde\alpha_{-n}^i+ \sum_jS_{ij}\alpha_n^j
  )\kket{h;S, \eta} &= 0
  \qquad\qquad\qquad\qquad
  S_{ij} = v^{\mu *}_{i}S_{\mu\nu}v^\nu_{j}
  \\
  ( \tilde\alpha_{-n}^{i\dagger}+
  \sum_jS_{ij}^*\alpha_n^{j\dagger}
  )\kket{h;S, \eta} &= 0
  \qquad\qquad\qquad\qquad
  S_{ij}^* = v^{\mu}_{i}S_{\mu\nu}v^{\nu *}_{j} \,.
  \end{split}
  \label{Sdef}
\end{equation}
$S_{ij}$ is unitary because $S^{\mu \nu}$ is orthogonal, moreover
$S_{ij}=0$ if $\alpha_i\neq\alpha_j$ because $S$ and $A$ commute.
Using this one finds the solution to the equations above:
\begin{align}
  \kket{h;S}_{bosonic} &=
  \exp\left( 
    \sum_{\substack{ij\\n<0}}\frac{2}{\frac{\alpha_i}{2\pi}+n} 
    \tilde\alpha^{i\dagger}_{-n} S_{ij}\alpha^{j}_{n} -
    \sum_{\substack{ij\\n\geq 0}}\frac{2}{\frac{\alpha_i}{2\pi}+n} 
    \tilde\alpha^{i}_{-n}S^*_{ij}\alpha^{j\dagger}_{n}
  \right) \ket{h} \,.
  \label{ishibashisol}
\end{align}
This solution is unique up to a constant, more precisely one can
multiply this state by any operator commuting with $X^\mu$, this is
exactly what will happen when the fermionic part is included.

The fermionic fields satisfy
\begin{align*}
  \psi_R^\mu(\sigma +2\pi , \tau) &= - A_R^{\mu \nu}
  \psi_R^\nu(\sigma,\tau)
  &
  \psi_L^\mu(\sigma +2\pi , \tau) &= - A_L^{\mu \nu}
  \psi_L^\nu(\sigma,\tau) \,.
\end{align*}
The mode expansions are
\begin{align*}
  \psi_R^\mu(\sigma,\tau) &= 
  \sum_{n,i} \psi^i_{n} f^{\mu}_{R,i,n}(\sigma - \tau) 
  + \psi^{i\dagger}_{n} f^{\mu *}_{R,i,n}(\sigma - \tau) 
  \\
  \psi_L^\mu(\sigma,\tau) &= 
  \sum_{n,i} \tilde\psi^i_{n} f^{\mu}_{L,i,n}(\sigma + \tau) 
  + \tilde\psi^{i\dagger}_{n} f^{\mu *}_{L,i,n}(\sigma + \tau) \,.
\end{align*}
In the sum $n$ runs through the integers and $i=1,2,3,4$, and the
functions $f_{R,i,n}^\mu,f_{L,i,n}^\mu$ correspond to $-A_R$ and
$-A_L$ respectively and are defined as in the bosonic case.  The
anticommutation relations are
\begin{align*}
  \{ \psi^i_n , \psi^{j\dagger}_{m} \} &= 
  \delta^{ij} \delta_{nm} 
  &
  \{ \tilde\psi^i_{n} , \tilde\psi^{j\dagger}_{m} \} &=
  \delta^{ij} \delta_{nm} \,,
\end{align*}
all other anticommutators are zero. We have to specify which operators
are annihilation operators and which are creation operators.  For
nonzero modes the ones that raise the energy are creation operators,
for the zero modes there is a choice. The adjoint of a creation
operator is an annihilation operator.  We will take $\psi^i_{n \geq
  0}$, $\tilde\psi^{i\dagger}_{n \geq 0}$, $\psi^{i\dagger}_{n<0}$ and
$\tilde\psi^i_{n <0}$ to be annihilation operators and the rest to be
creation operators. The point is that the complex structure of $\RR^8$ 
gives a natural splitting of the zero modes into annihilation and 
creation operators.    

We want to solve the gluing condition
\begin{align*}
  ( \psi^\mu_L(\tau=0,\sigma) 
  + i \eta S^{\mu\nu}\psi^\nu_R(\tau=0,\sigma))   
  \kket{h;S,\eta} &= 0 \,,  
\end{align*}
in a twisted sector with $A_L=A_R$. Expanding this into modes as in
the bosonic case one gets
\begin{align*}
  (\tilde\psi^i_n + i \eta \sum_jS_{ij} \psi^j_{n})
  \kket{h;S,\eta} &= 0 
  &
  (\tilde\psi^{i\dagger}_{n} + 
  i \eta \sum_jS^*_{ij} \psi^{j\dagger}_{n})
  \kket{h;S,\eta} &= 0
\end{align*}
where $S_{ij}$ is the unitary matrix defined in \myref{Sdef}.  The
solution is now readily found to be
\begin{align*}
  \kket{h;S,\eta}_{fermionic} = 
  \exp(-i \eta \sum_{\substack{ij \\ n <0}} \tilde\psi^{i\dagger}_{n}
  S_{ij} \psi^j_{n} - i \eta \sum_{\substack{ij \\ n \geq 0}}
  \tilde\psi^i_{n} S^*_{ij} \psi^{j\dagger}_{n} )
  \ket{h}
\end{align*}
where $\ket{h}$ is annihilated by all the annihilation operators.
$\ket{h}$ is unique up to a constant.

Let us now recapitulate. Gluing conditions are characterized by
$S^{\mu\nu}, \eta$.  In a sector twisted by $h \in \Gamma$ there is a
state solving the gluing condition if $[A(h) , S]=0$ and
$A_L(h)=A_R(h)$. The solution is unique up to a multiplicative
constant and its explicit form is
\begin{multline}
  \kket{h;S,\eta} = 
  \kket{h;S,\eta}_{fermionic} \kket{h;S}_{bosonic} 
  \\
  = 
  \exp\left({
    -i \eta \sum_{\substack{ij \\ n <0}} \tilde\psi^{i\dagger}_{n}
    S_{ij} \psi^j_{n} -i\eta  \sum_{\substack{ij \\ n \geq 0}}
    \tilde\psi^i_{n} S^*_{ij} \psi^{j\dagger}_{n} 
  }\right)
  \\ 
  \times
  \exp\left({
    \sum_{\substack{ij \\ n < 0}}\frac{2}{\frac{\alpha_i}{2\pi}+n} 
    \tilde\alpha^{i\dagger}_{-n} S_{ij}\alpha^{j}_{n} -
    \sum_{\substack{ij \\ n \geq 0}}\frac{2}{\frac{\alpha_i}{2\pi}+n} 
    \tilde\alpha^{i}_{-n}S^*_{ij}\alpha^{j\dagger}_{n}
  }\right) 
  \,\ket{h} \,.
\end{multline}
We will assume that $\ket{h}$ is normalized to have norm 1.

\subsection{$\Gamma$-invariant boundary states}
\label{Boundaryprojected}

In the previous section we found the boundary states on the covering
space, i.e. without imposing $\Gamma$ invariance.  Now we will discuss
how to find the $\Gamma$-invariant states.  The first thing to note is
that the gluing condition is not $\Gamma$-invariant; from
\myref{orbihilbert} and \myref{ishibashiclosed} we see that if a state
$\ket{a}$ satisfies the gluing condition with parameters $S , \eta$
and $g \in \Gamma$ then $g \ket{a}$ satisfies the gluing condition
with parameters $^gS,\, ^g\eta$ where
\begin{align*}
  ^gS &= A(g) S A^{-1}(g)  
  &
  \,^g\eta &= \mbox{sign}(A_L(g)A_R^{-1}(g)) \eta \,.
\end{align*}
Here $A_L(g)A_R^{-1}(g)$ is either the identity or minus the identity
and $\mbox{sign}(A_L(g)A_R^{-1}(g))$ should be understood as $1$ or
$-1$ respectively.  In other words $\eta$ is unchanged if $A_L(g) =
A_R(g)$ and changes sign otherwise. The following identities hold:
\begin{align*}
  ^{(g_1g_2)}S&=\,^{g_1}(^{g_2}S) &
  ^{(g_1g_2)}\eta&=\,^{g_1}(^{g_2}\eta)\,. 
\end{align*}
If a state $\ket{h}$ is twisted by $h \in \Gamma$ then $g \ket{h}$ is
twisted by $g^{-1} h g$.  Now let us look at the action of $g \in
\Gamma$ on the boundary state $\kket{h;S,\eta}$.  The solution to a
gluing condition is unique up to a constant therefore 
\begin{align}
  g\kket{h;S, \eta} = \phi(h,g,S,\eta) \kket{ghg^{-1}; \,^gS\,, 
     ^g\eta} \,,
  \label{konst}
\end{align}
where $\phi(h,g,S,\eta)$ is a constant.  Strictly speaking there is
also a possible momentum ambiguity in the solution, but we set
momentum equal to zero and the action of $g$ does not change the
momentum away from zero.  Furthermore both states above have the form
of an exponential acting on a lowest weight state. In the sector
twisted by $h$ we expanded the fields $X^\mu$ and $\psi^\mu$ in terms
of creation and annihilation operators. If $\gamma$ is an annihilation
operator in the expansion then $g \gamma g^{-1}$ is also an
annihilation operator and similarly for creation operators. This
important point follows from the fact that $g$ maps the functions
$f_{i,n}$ defined in \myref{ffunctions} into linear combinations of
the corresponding $f_{j,n}$ in the sector twisted by $ghg^{-1}$. $n$
is kept fixed but there is a possible linear combination in $j$. Here
it is essential that the orbifold group $\Gamma_0$ is in $U(4)$.  This
important point means that the ground state $\ket{h}$, is mapped into
the ground state $\ket{ghg^{-1}}$ up to a constant, and excited states
are mapped into excited states. In \myref{konst} we can thus fix the
constants, $\phi(h,g,S,\eta)$, by comparing the coefficients of the
ground state which do not depend on $S$ and $\eta$:
\begin{align}
  g\ket{h} &= \phi(h,g) \ket{ghg^{-1}}
 \label{defphi}
 \\
 g\kket{h;S, \eta} &= \phi(h,g)\kket{ghg^{-1};\,^gS\,,\,^g\eta} \,.
 \label{konst1} 
\end{align}
Here $g \in \Gamma$ and $h \in \Gamma_{sym}$ because the twists in a
boundary state are always symmetric as explained earlier. In order to
know the full string theory one also needs the action of $\Gamma$ in
asymmetric twisted sectors, but that is irrelevant for us.  We note
that the exponentials are mapped into each other thus the phase only
comes from the ground state.  $\phi(h,g)$ has unit norm since $g$ is a
unitary operator.  These constants are still convention dependent to a
certain degree as there is a possibility of redefining the states
$\ket{h}$ up to a phase.  Partially fixing these constants will be the
subject of the next section.  The $\phi(h,g)$ depends on the exact
definition of the orbifold and they change when discrete torsion is
introduced.  From \myref{defphi} it follows that
\begin{align}
  \phi(h, g_1g_2) = \phi(h,g_2)\phi(g_2hg_2^{-1},g_1) \,.
  \label{multphi}
\end{align}

To get a $\Gamma$-invariant state we thus need to sum over an
equivalence class of gluing conditions and an equivalence class of
twists. We will now find the $\Gamma$-invariant states for each
equivalence class of gluing conditions and an equivalence class of
twists; these will then constitute a basis for all $\Gamma$-invariant
states.  Consider the gluing condition $S, \eta$ and the twist $h \in
\Gamma$.  The state $\kket{h;S, \eta}$ is not $\Gamma$-invariant, we
need to project it onto the space of $\Gamma$-invariant states with
the projector
\begin{align*}
  P = \frac1{|\Gamma|} \sum_{g \in \Gamma} g \,.
\end{align*}
An important subgroup of $\Gamma$ is the group
$\Gamma_{h;S}\subset\Gamma_{sym}$ which leaves $S,\eta$ and $h$
invariant. It does not depend on whether $\eta=1$ or $\eta=-1$.  Now
\begin{align*}
  P \kket{h;S, \eta} &= 
  \frac1{|\Gamma|} \sum_{g \in \Gamma} g \kket{h;S, \eta}  = 
  \frac1{|\Gamma|}
  \sum_{[f] \in \Gamma / \Gamma_{h;S}} 
  \sum_{k \in \Gamma_{h;S}} f k \kket{h;S, \eta} \,.
\end{align*}
Here $f$ is a representative of the class $[f] \in \Gamma /
\Gamma_{h;S}$.  The states $f k \kket{h;S, \eta}$ for different
$[f]$'s are images of each other and are linearly independent since
they have different twists or solve different gluing conditions or
both.  The important part is thus
\begin{align*}
  \sum_{k \in \Gamma_{h;S}} k \kket{h;S, \eta} = \sum_{k \in
    \Gamma_{h;S}} \phi(k,h) \kket{h;S, \eta} \,.
\end{align*}

\subsection{Partial fixing of $\phi(g,h)$}
\label{Partialfixing}

Now we will use the the assumption that all the fractional
D-instantons exist in one of the orbifold theories, which we we call
the theory without discrete torsion.  By all the fractional instantons
we mean that there is one instanton and one anti instanton for each
representation of $\Gamma_0$.  For D-instantons $S^{\mu \nu} =1$, so
we consider states of the form $\kket{h,1,\eta}$. $\eta=1$ and
$\eta=-1$ are equivalent in the sense described above: they are mapped
into each other by $(-1)^{F_L}$ for instance. The twist $h$ is
composed of a part in $\Gamma_0$ and another part which is NSNS or RR.
$h_1$ and $h_2$ are equivalent if they are both RR or both NSNS and
their $\Gamma_0$ part are conjugate in $\Gamma_0$. The number of
equivalence classes of $\Gamma_0$ is the same as the number of
irreducible representations of $\Gamma_0$.  We thus see that the
number of equivalence classes of boundary states is exactly equal to
the number of fractional branes expected.  We thus conclude that they
all have to survive the $\Gamma_{h,1}$ projection:
\begin{align*}
  k \kket{h,1, \eta} =    \kket{h,1, \eta} 
  \qquad\qquad\qquad
  k \in \Gamma_{h,1}\,,
\end{align*}    
that is $\phi(k,h)=1$, or $k \ket{h} = \ket{h}$ for $h \in
\Gamma_{sym}$ and $k \in \Gamma_{h,1}$.  Now what is the structure of
$\Gamma_{h,1}$?  $k$ is in $\Gamma_{h,1}$ if and only if the
$\Gamma_0$ part of $k$ and $h$ commute and $k \in \Gamma_{sym}$. The
element $(-1)^F =(-1)^{F_L}(-1)^{F_R}$ is an example of this thus
$(-1)^F \ket{h} = \ket{h}$ for all $h \in \Gamma_{sym}$.  This is
actually the requirement that the theory is type IIB.  This is as
expected since type IIA does not have all these D-instantons. Type IIA
only has one kind of instanton, no anti instantons.  We thus conclude
that if $h$ and $k$ are commuting elements of $\Gamma_{sym}$
\begin{align}
  k \ket{h} = \ket{h}
  \qquad\qquad\mbox{if}\qquad hk=kh  \,.
  \label{equal}
\end{align}
When they do not commute we use the phase ambiguity in the definition
of the lowest weight states to require
\begin{align} 
  k \ket{h} = \ket{khk^{-1}} \,.
  \label{equalgen}
\end{align} 
This assignment of phases is possible exactly because of
\myref{equal}. We thus conclude that $\phi(h,k)=1$ for $h,k \in
\Gamma_{sym}$. The only undetermined part is then when $k$ is not
symmetric. Then $k = k_0 (-1)^{F_L}$ where $k_0 \in \Gamma_{sym}$.
From \myref{multphi} we get
\begin{align*}
  \phi(h,k_0 (-1)^{F_L})= \phi(h, (-1)^{F_L})\phi(h,k_0)
  = \phi(h, (-1)^{F_L}) \,.
\end{align*}
Therefore the only undetermined part is $\phi(h, (-1)^{F_L})$.
Similarly we easily derive from \myref{multphi} that $\phi(h,
(-1)^{F_L}) = \phi(k^{-1}hk,(-1)^{F_L})$, for any $k \in
\Gamma_{sym}$, i.e.  $\phi(h, (-1)^{F_L})$ for $h \in \Gamma_{sym}$ is
a class function in $\Gamma_{sym}$.  Furthermore we know that the
tachyonic NSNS ground state $\ket{1}$ is odd under $(-1)^{F_L}$, hence
$\phi(1, (-1)^{F_L})=-1$.

In summary:
\begin{align}
  g\kket{h;S, \eta} = \phi(h,g) \kket{ghg^{-1}; \,^gS,\,^g\eta} \,,
  \label{actiononbdy}
\end{align}
$\phi(h,g)=1$ whenever $g \in \Gamma_{sym}$. The only indeterminacy in
$\phi$ at the moment is given by $\phi(h, (-1)^{F_L})$ which is a
class function in $h \in \Gamma_{sym}$. Furthermore we know that
$\phi(1, (-1)^{F_L})=-1$.  
Introducing discrete torsion will modify $\phi$ by multiplication by
$\varepsilon$.

\subsection{The most general $\Gamma$ invariant boundary state}
\label{Boundarygeneral}

After fixing $\phi(h,k)=1$ in the relevant cases we can write down the
form of the most general $\Gamma$ invariant boundary state. 

In order to find the physical states it is necessary to use linear
combinations of the boundary states $\kket{h;S,\eta}$ for
$h\in\Gamma_{S}$ whereas we do not need to sum over the gluing
condition $S,\eta$ which would lead to a superposition of branes of
different dimensionality and orientation. While it is certainly
possible to superpose these states it does not add new information to
our analysis.

For a given gluing condition $S,\eta$ a generic boundary state in the
covering space is a linear combination of the boundary states found in
each twisted sector corresponding to the elements of $\Gamma_{S}$:
\begin{align*}
  &\kket{c\,;S,\eta}_{unprojected} = 
  \frac{\sqrt{|\Gamma|}}{|\Gamma_S|}
  \sum_{h\in\Gamma_{S}}c(h)\kket{h;S,\eta} \,.
\end{align*} 
The normalization of the coefficients $c(h)$ is chosen for later
convenience. In the orbifolded theory only the $\Gamma$-invariant part
survives:
\begin{align*} 
  &\kket{c\,;S,\eta} = 
  \frac{1}{\sqrt{|\Gamma|}|\Gamma_S|} \sum_{g\in\Gamma}g
  \sum_{h\in\Gamma_{S}}\varepsilon(h,g)c(h)\kket{h;S,\eta} \,.
\intertext{Let us now split the sum over the elements of $\Gamma$ in
  the following way:}
  &\kket{c\,;S,\eta} = 
  \frac{1}{\sqrt{|\Gamma|}|\Gamma_S|} 
  \sum_{[f]\in\Gamma/\Gamma_{S}} f
  \sum_{k\in\Gamma_{S}}
  \sum_{h\in\Gamma_{S}}\varepsilon(khk^{-1},f)
  \varepsilon(h,k)c(h)k\kket{h;S,\eta}  \,. 
\intertext{Using $\phi(h,k)=1$ for $k \in \Gamma_{S} \subset
  \Gamma_{sym}$ and changing variables within the last two sums}
  &\kket{c\,;S,\eta} = 
  \frac{1}{\sqrt{|\Gamma|}|\Gamma_S|} 
  \sum_{[f]\in\Gamma/\Gamma_{S}} f
  \sum_{h\in\Gamma_{S}}   \varepsilon(h,f)
  \left(\sum_{k\in\Gamma_{S}}
  c(k^{-1}hk)\varepsilon(k^{-1}hk,k)\right)\kket{h;S,\eta} \,.
\end{align*}
The sum inside the parenthesis acts as a projection operator onto the
space of functions $f$ satisfying $f(khk^{-1})=f(h)\varepsilon(h,k)$.
Therefore we can without loss of generality restrict $c(h)$ to be such
a function i.e.
\begin{align}
  c(khk^{-1}) &= c(h)\varepsilon(h,k)
  &
  h,k\in\Gamma_{S}\,.
  \label{ctorsion}
\end{align}
Note that $(-1)^{F_L}$ which flips the sign of $\eta$ always appears
in the sum over $f$. The projected state therefore depends on $\eta$
at most through an overall sign factor and we can make a choice of
$\eta=+1$ to start with.  We thus conclude that the most general
$\Gamma$-invariant boundary state is
\begin{align*}
  \kket{c\,;S} &= 
  \frac{1}{\sqrt{|\Gamma|}} 
  \sum_{[f]\in\Gamma/\Gamma_{S}} f
  \sum_{h\in\Gamma_{S}}\varepsilon(h,f)c(h)\kket{h;S,+} \,.
\end{align*}
with $c(h)$ a function satisfying \myref{ctorsion}.  It shows in
particular that $c(h)$ vanishes unless $\varepsilon(h,k)=1$ for all $k
\in \Gamma_{h,S}$.

In the next section we will impose open-closed string duality on the
states to discover which ones are physical, e.g.  half a D-brane is
not physical.

\section{Physical boundary states from open-closed duality}
\label{Openclosed}

Not all $\Gamma$-invariant boundary states are actually realized in
the spectrum. The condition which fixes the set of physical states is
open string closed string duality which is the principle that a closed
string propagator between two boundary states has a dual description
in terms of an open string partition function \cite{Cardy:1989ir}:

\EPSFIGURE[!h]{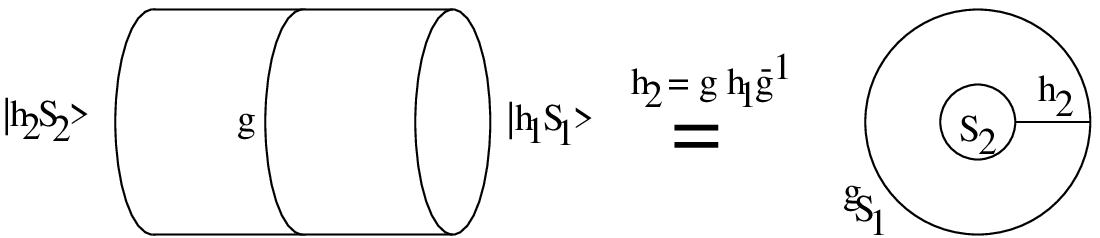}{ The dual descriptions of the cylinder
  amplitude.  In the closed string picture the states are prepared in
  the twisted sectors $h_{1,2}$ and the operator $g$ is inserted in
  the propagator.  In the open string picture $g$ identifies the
  sector of the open string theory and $h_2$ gets inserted in the
  propagator.}

Using $P^\dagger P=P$ we can write the closed string propagator as
\begin{align}
  \nn
  &\bbra{c_2;S_2} e^{-\frac{2\pi}{l}H} \kket{c_1;S_1} =
  \\ \nn &\quad=
  \frac{1}{|\Gamma_{S_1}||\Gamma_{S_2}|}
  \sum_{g\in\Gamma}
  \sum_{\substack{h_2\in\Gamma_{S_2} \\ h_1\in\Gamma_{S_1}}}
  \bbra{h_2;S_2,+}
  c_2^*(h_2) g\varepsilon(h_1,g)
  c_1(h_1)e^{-\frac{2\pi}{l}H}\kket{h_1;S_1,+} 
  \\ \nn &\quad= 
  \frac{1}{|\Gamma_{S_1}||\Gamma_{S_2}|} 
  \sum_{\substack{[f]\in\Gamma_{S_2}\backslash\Gamma/\Gamma_{S_1} 
      \\ k_1\in\Gamma_{S_1}
      \\ k_2\in\Gamma_{S_2}}}
  \frac{1}{|\Gamma_{S_2}\cap\Gamma_{\,^fS_1}|} \times
  \\ &\qquad\qquad\times
  \sum_{\substack{h_2\in\Gamma_{S_2} \\ h_1\in\Gamma_{S_1}}}
  \bbra{h_2;S_2,+}
  c_2^*(h_2) k_2^{-1}fk_1\varepsilon(h_1,k_2^{-1}fk_1)
  c_1(h_1)e^{-\frac{2\pi}{l}H}\kket{h_1;S_1,+} \,.
\intertext{Here we wrote $g=k_2^{-1}fk_1$ with $k_i\in\Gamma_{S_i}$
  and $f$ an arbitrary representative of the coset
  $\Gamma_{S_2}\backslash\Gamma/\Gamma_{S_1}$ and we compensated for
  the overcounting by dividing by the order of
  $\Gamma_{\,^fS_1}\cap\Gamma_{S_2}$.  After the change of variables 
  $k_2h_2k_2^{-1}\longrightarrow h_2$ and
  $k_1fh_1f^{-1}k_1^{-1}\longrightarrow h_2$ and using
  \myref{actiononbdy} we obtain}  
  \nn
  &\bbra{c_2;S_2} e^{-\frac{2\pi}{l}H} \kket{c_1;S_1} =
  \\ \nn&\quad= 
  \frac{1}{|\Gamma_{S_1}||\Gamma_{S_2}|} 
  \sum_{\substack{[f]\in\Gamma_{S_2}\backslash\Gamma/\Gamma_{S_1} 
      \\ k_1\in\Gamma_{S_1}
      \\ k_2\in\Gamma_{S_2}}}
  \frac{1}{|\Gamma_{S_2}\cap\Gamma_{\,^fS_1}|} \times
  \\ \nn&\qquad\times
  \sum_{\substack{h_2\in\Gamma_{S_2} \\ h_1\in\Gamma_{\,^fS_1}}}
  c_2^*(k_2^{-1}h_2k_2) \varepsilon(k_1^{-1}f^{-1}h_1fk_1,k_2^{-1}fk_1)
  c_1(k_1^{-1}f^{-1}h_1fk_1)\phi(h_1,f)
  \times  \\ &\qquad\times
  \bbra{h_2;S_2,+}e^{-\frac{2\pi}{l}H}
  \kket{h_1;\,^fS_1,\,^f+} \,.
  \label{sandwich}
\end{align}

To proceed we need to express the general closed string scalar product
in terms of an open string trace: 
\begin{align*}
  \bbra{h_2;S_2,\eta_2}e^{-
    \frac{2\pi}{l}H}\kket{h_1,S_1,\eta_1} &=  \delta_{h_1h_2}
  \mbox{Tr}_{\substack{S_1\eta_1 \\ S_2(-\eta_2)}}
  (h_1e^{-\pi lH_{o}}) \,,
\end{align*}
where by $\mbox{Tr}_{\substack{S_1\eta_1 \\ S_2\eta_2}}$ we mean the
trace over the open string Hilbert space with boundary conditions
$S_{i=1,2},\eta_{i=1,2}$ at the two ends of the worldsheet.  As
discussed, the above boundary states have a phase ambiguity which
however drops out in expressions involving the sandwiched vectors. A
priori there could be an ambiguity in the action of $h_1$ in the open
string spectrum but since the left hand side of the above equation is
completely well-defined, it defines $h_1$. This formula follows from
the fact that the path integral over the worldsheet with boundaries
can be calculated in two channels.  Note that the bra in the
propagator satisfies the conjugate gluing condition which is why the
sign of $\eta_2$ is reversed.  Also note that the untwisted closed
sector, $h_1=h_2=1$ is the NSNS sector.

Going back to \myref{sandwich} we can now express the closed string
propagator in terms of open string traces.  Using \myref{ctorsion} and 
\myref{disrep} the coefficient in the sum simplifies for $h=h_1=h_2$
as 
\begin{align*}
  c_2^*(k_2^{-1}hk_2) \varepsilon(k_1^{-1}f^{-1}hfk_1,k_2^{-1}fk_1)
  c_1(k_1^{-1}f^{-1}hfk_1) = 
  c_2^*(h)c_1(f^{-1}hf)\varepsilon(f^{-1}hf,f)\,,
\end{align*}
and as $k_{1,2}$ drop out of the summand this leads to
\begin{multline}
  \label{finalformula}  
  \bbra{c_2;S_2} e^{-\frac{2\pi}{l}H} \kket{c_1;S_1} =
  \sum_{[f]\in\Gamma_{S_2}\backslash\Gamma/\Gamma_{S_1}}
  \frac{1}
  {|\Gamma_{S_2}\cap\Gamma_{\,^fS_1}|}\times
  \\ \times 
  \sum_{h\in\Gamma_{S_2}\cap\Gamma_{\,^fS_1}}
  c_2^*(h) \varepsilon(f^{-1}hf,f) c_1(f^{-1}hf)
  \phi(h,f)
  \mbox{Tr}_{\substack{\,^fS_1,\,^f+ \\ S_2,-}}
  (he^{-\pi lH_{o}}) \,.
\end{multline}
Having derived the final formula for the partition function in the
open string channel we will turn to a discussion of it's consequences.

The sum over $[f]\in\Gamma_{S_2}\backslash\Gamma/\Gamma_{S_1}$ is a
sum over the different sectors of the open string while the sum over
$h\in\Gamma_{S_2}\cap\Gamma_{\,^fS_1}$ projects out some of the states
in the sector twisted by $f$. The physical condition is that the
an integer number of states should be picked out, or more precisely
that the projection selects every irreducible representation a
nonnegative number of times.  In general such a projector has the form 
\begin{align*}
  P = \frac{1}{|G|}\sum_{g \in G} \chi(g)g \,,
\end{align*}
where $\chi$ is a character of the group $G$, i.e.
$\chi(g)=\mbox{Tr}_R(g)$ for some representation $R$.  There is one
more issue to understand, namely the overall sign that a sector
appears with in the partition function.  This is the standard
partition function of string theory which -- from the D-brane
worldvolume's point of view -- has an $(-1)^F_{worldvolume}$ inserted.
In other words worldvolume bosons count positively and worldvolume
fermions count negatively.  In the above sum over open string sectors
the bosons come from the NS sector, which has antiperiodic boundary
conditions on the worldsheet fermions. This happens for $f \in
\Gamma_{sym}$ since $^f+ = +$ for $f \in \Gamma_{sym}$.  Similarly the
states are worldvolume fermions in the R sector.  The R sector is for
asymmetric $f$, i.e. of the form $f = (-1)^{F_L}f_0$ for $f_0 \in
\Gamma_{sym}$.  We also note that the constant
$|\Gamma_{S_2}\cap\Gamma_{\,^fS_1}|$ in \myref{finalformula} is
exactly the order of the group which is being summed over.

Combining everything we are ready to state the conditions on the
coefficients in \myref{finalformula}.  For worldvolume bosons, $f \in
\Gamma_{sym}$
\begin{align*}
  c_2^*(h)  c_1(f^{-1}hf)\varepsilon(f^{-1}hf,f)\,,
\end{align*}
is a character of $\Gamma_{S_2}\cap\Gamma_{\,^fS_1}$.  Here we used
that $\phi(h,f)=1$ for $f$ symmetric.

For worldvolume fermions, $f=(-1)^{F_L} f_0$, $f_0 \in \Gamma_{sym}$
\begin{multline}
  c_2^*(h)  c_1(f^{-1}hf)\varepsilon(f^{-1}hf,f) \phi(h,(-1)^{F_L})= 
  \\
  =   c_2^*(h)  c_1(f_0^{-1}hf_0)\varepsilon(f_0^{-1}hf_0,f_0) 
     \varepsilon(h,(-1)^{F_L}) \phi(h,(-1)^{F_L})\,,
\end{multline}
is minus a character of $\Gamma_{S_2}\cap\Gamma_{\,^fS_1}$.  Here we
used that $\phi(h,f)=\phi(h,(-1)^{F_L} f_0) = \phi(h,(-1)^{F_L})$. It
is minus a character because worldvolume fermions are counted with a
minus sign.

Let us discuss the compatibility of these two conditions.  They are
compatible if and only if $h \rightarrow \phi(h,(-1)^{F_L})$ is minus
a character. Here we used that $\varepsilon(h,(-1)^{F_L})$ is a
character which has an inverse character.  Since we believe in the
consistency of the theory we thus conclude that $\phi(h,(-1)^{F_L})$
is minus a character. So far the only knowledge we had about
$\phi(h,(-1)^{F_L})$ was that it takes the values $1$ or $-1$, it was
a class function and that $\phi(1,(-1)^{F_L})=-1$. These facts agree
nicely with the conclusion that it is minus a character. Later we will
have more to say about it.
% Furthermore $\varepsilon(h,f)$ is an invertible character, so it drops
% out of the condition.  

We thus conclude that the physical D-brane states are characterized by
a gluing condition $S$, a function $c(h)$, $h \in \Gamma_S$ which
satisfies
\begin{align}
  &c(khk^{-1}) = c(h)\varepsilon(h,k) 
  \qquad\qquad\qquad
  h,k \in \Gamma_S\,,
  \label{condone}
\intertext{and for any two D-branes characterized by $S_1,S_2$ and
  $c_1 , c_2$,}
  &c_2^*(h) c_1(f^{-1}hf)\varepsilon(f^{-1}hf,f)\,,
  \label{condtwo}
\end{align}
is a character of $\Gamma_{S_2}\cap\Gamma_{\,^fS_1}$ for any $f \in
\Gamma_{sym}$.  The set of allowed states constitute a cone, i.e. it
is additive and an allowed state can be multiplied by a nonnegative
integer.

In the next two sections we will discuss what the allowed set of
D-brane states are. Starting with the case without discrete torsion
followed by the case with discrete torsion.

\subsection{No discrete torsion}
\label{Nodiscretetorsion}

In case of no discrete torsion the coefficients $c$ are class
functions.  The physical condition implies that the function $c_2^*(h)
c_1(f^{-1}hf)$ is a character of $\Gamma_{S_2}\cap\Gamma_{\,^fS_1}$
for any $f \in \Gamma_{sym}$.  In order to see what this means for the
individual functions $c(h)$ we shall now consider specific examples:
\paragraph{D-instantons} 
First let's look at the product of boundary states with $S_1= S_2=
\unit$. These gluing conditions are left invariant by $\Gamma_{S_1} =
\Gamma_{S_2} = \Gamma_{sym}$. The only $f$ to consider is $f=1$.  and
$\Gamma_{\,^fS_1} = \Gamma_{sym}$.  We thus need $c_2^*(h) c_1(h)$ to
be a character of $\Gamma_{sym}$.  This is satisfied if the $c(h)$ are
taken to be the characters of $\Gamma_{sym}$. This is exactly our
assumption through the paper that all the fractional D-instantons
exist.  We remember that characters are of the form $c(h)= Tr_R(h)$
for some representation $R$ of $\Gamma_{sym}$.  Of these the
elementary ones correspond to the irreducible representations.
\paragraph{D-instanton with another boundary state}
We now let $S_2$ be generic while restricting $S_1=\unit$.  The only
$f$ to check is again $f=1$.  Again $\Gamma_{S_1} = \Gamma_{sym}$ so
the physical condition is that $c_2^*(h) c_1(h)$ is a character of
$\Gamma_{S_2}$.  Since we can especially choose $c_1(h)=1$ it implies
that $c_2^*(h)$ and hence $c_2(h)$ is a character.  It is natural to
call those D-branes elementary for which this representation is
irreducible.
\paragraph{Two arbitrary D-branes}
Having found that consistent interaction with the elementary
D-instanton requires the coefficients of any boundary state to be
characters, we now show that the interaction of any two of these is
consistent with open-closed duality.  To this end note that $f \in
\Gamma_{sym}$ generates a group homomorphism:
\begin{align*}
  f&:\Gamma_{\,^fS_1} \longmapsto \Gamma_{S_1} &
  f&: h \longmapsto f^{-1}hf \,,
\end{align*}
implying that if $c_1$ is a character of $\Gamma_{S_1}$ then $h\mapsto
c_1(f^{-1}hf)$ is a character of $\Gamma_{\,^fS_1}$. Furthermore when
a character is restricted to a subgroup then it becomes a character of
that subgroup so that both $h\mapsto c_2(h)$ and $h\mapsto
c_1(f^{-1}hf)$ are characters of $\Gamma_{S_2}\cap\Gamma_{\,^fS_1}$.
Since the product of characters is again a character, the condition
for the consistency with open-closed duality is satisfied.

\subsection{Discrete torsion}
\label{Yesdiscretetorsion}

Now we turn to the general case. It was conjectured by Douglas and
Fiol \cite{Douglas:1998xa,Douglas:1999hq} that in this case the
D-branes are classified by projective representations.  Remembering
the connection between discrete torsion and projective representations
it is a natural generalization of the ordinary case. Let us show that
this indeed works.  For a given gluing condition $S$ we need a
function $c(h)$ defined on $\Gamma_S$. Let $\gamma : \Gamma_S
\rightarrow GL(n, \CC)$ be a projective representation of $\Gamma_S$
corresponding to the given $\varepsilon$. This means that $\gamma$
satisfies \myref{projdef} and $c(g,h)$ is related to $\varepsilon$ by
\myref{cepsilon}.  Let us now define
\begin{align}
  c(h) &= \mbox{Tr}(\gamma(h))  & h \in \Gamma_S\,.
  \label{charproj}
\end{align}
Our claim is now that this set of coefficients $c$ will span the set
of physical states.  We have to check two conditions. One is the
condition \myref{condone} and the other is the physical condition
\myref{condtwo}.  We first prove \myref{condone}:  
\begin{align*}
  c(khk^{-1}) &= \mbox{Tr}(\gamma(khk^{-1}) 
  = \frac{c(k^{-1},k)c(h,1)}{c(kh,k^{-1})c(k,h)} 
  \mbox{Tr}(\gamma(h))\,.
\intertext{Here we used \myref{projdef}, the cyclicity of the trace
  and \myref{projdef} again.  Now substitute the definition of
  $\varepsilon$ and apply \myref{projdef} once more:}
  c(khk^{-1})&= \varepsilon(h,k)c(h)
  \frac{c(k^{-1},k)c(h,1)}{c(kh,k^{-1})c(khk^{-1},k)} 
  = \varepsilon(h,k)c(h)\frac{c(h,1)}{c(kh,1)} 
  = \varepsilon(h,k)c(h)\,,
\end{align*} 
exactly as desired.  The physical condition \myref{condtwo} is a bit
harder.  With notation as above we need to show that
$\varepsilon(f^{-1}hf,f)c_1(f^{-1}hf)c_2^*(h)$ is a character.  We
have
\begin{align*}
  \varepsilon(f^{-1}hf,f)c_1(f^{-1}hf)c_2^*(h) &=
  \varepsilon(f^{-1}hf,f)
  \mbox{Tr}(\gamma_1(f^{-1}hf))\mbox{Tr}(\gamma_2(h))^*  
  \\ &= \mbox{Tr} (\varepsilon(f^{-1}hf,f)
  \gamma_1(f^{-1}hf)\otimes \gamma^*_2(h))\,,
\end{align*}
where the last trace is over the tensor product representation.  This
is a character of $\Gamma_{S_2}\cap\Gamma_{\,^fS_1}$ if $R(h)=
\varepsilon(f^{-1}hf,f)\gamma_1(f^{-1}hf)\otimes \gamma^*_2(h)$ is a
proper representation, i.e. not projective. It indeed is as we will
now show.
\begin{align*}
  R(g)R(h) &= 
  \varepsilon(f^{-1}gf,f)\varepsilon(f^{-1}hf,f)
  (\gamma_1(f^{-1}gf)\gamma_1(f^{-1}hf)
  \otimes 
  (\gamma^*_2(g) \gamma^*_2(h)) 
  \\ &= \frac{\varepsilon(f^{-1}gf,f)\varepsilon(f^{-1}hf,f)}
  {\varepsilon(f^{-1}ghf,f)}
  \frac{c(f^{-1}gf,f^{-1}hf)}{c(g,h)} R(gh) \,.
\end{align*}
We need to calculate the coefficient of $R(gh)$ and show that it is
one.  Plugging in the definition of $\varepsilon$ this coefficient
reads as
\begin{align*}
  \frac{c(g,f)c(h,f)c(f^{-1}gf,f^{-1}hf)c(f,f^{-1}gh,f)}
  {c(f,f^{-1}gf)c(f,f^{-1}hf)c(g,h)c(f,gh)} & =
  \frac{c(g,f)c(h,f)c(f,f^{-1}gf)c(gf,f^{-1}hf)}
  {c(f,f^{-1}gf)c(f,f^{-1}hf)c(g,hf)c(h,f)}
  \\ &= \frac{c(g,hf)c(f,f^{-1}hf)}
  {c(f,f^{-1}hf)c(g,hf)} = 1\,.
\end{align*}
Here we used \myref{projassoc}. We thus conclude that $R(g)R(h)=R(gh)$
as claimed.  Since $R$ is a representation its trace is a character.
We have thus proven that the physical condition is satisfied for the
traces of the projective representations.  Of course, the case without
discrete torsion is a special case of this.

\subsection{Determination of $\phi(h,(-1)^{F_L})$}
\label{Determinephi}

We see that the spectrum of open string states depends on the function
$\phi(h,(-1)^{F_L})$, with $h \in \Gamma_{sym}$.  Is there any way to
determine $\phi(h,(-1)^{F_L})$ ?  What we know about
$\phi(h,(-1)^{F_L})$ is that it takes the value $\pm 1$ and it is
minus a character.  It actually turns out that it could be any
character and the various choices differ by discrete torsion.  To see
that, write the orbifold group as $\Gamma = \{1,(-1)^{F_L}\} \times
\Gamma_{sym}$ and let $\chi(h)$ be any character of $\Gamma_{sym}$
that takes the values $\pm 1$. This particularly implies that $\chi =
\chi^{-1}$. Any element of $\Gamma$ can be written $((-1)^{F_L})^n h$,
where $n \in Z_2$ and $h \in \Gamma_{sym}$.  Now define a discrete
torsion as follows
\begin{align*}
  \varepsilon( ((-1)^{F_L})^{n_1} h_1, ((-1)^{F_L})^{n_2} h_2)
  = \chi(h_2)^{n_1} \chi(h_1)^{n_2} \,.
\end{align*}
It is easily seen that it satisfies the conditions \myref{disrep} and
\myref{modinv}.  Here it is essential that $\chi = \chi^{-1}$.  Now
introduction of this discrete torsion changes $\phi$ to
\begin{align*}
  \phi'(h, k) = \phi(h, k) \varepsilon(h,k) \,.
\end{align*}
Recall that $\phi(h,k)$ is defined for $h \in \Gamma_{sym}$ and $k \in
\Gamma$.  For $k \in \Gamma_{sym}$ we still have
\begin{align*}
  \phi'(h,k) = 1 \,,
\end{align*}
and
\begin{align*} 
  \phi'(h, (-1)^{F_L}) = \phi(h,(-1)^{F_L}) \varepsilon(h, (-1)^{F_L})
  = \phi(h,(-1)^{F_L}) \chi(h) \,.
\end{align*}
We thus see that $\phi'$ is still minus a character and it can be any
one taking the values $\pm 1$. Especially we could choose $\chi(h)$ to
be $- \phi(h,(-1)^{F_L})$ making $\phi'(h,(-1)^{F_L})=-1$.

\section{Conclusion}  
\label{Conclusion}
We have generalized the connection between the phases $\varepsilon$,
and projective representations to nonabelian groups \myref{cepsilon}.
We found explicit formulas for the boundary states with gluing
conditions in $U(4)$. Open-closed string duality is satisfied by
boundary states given as projective characters. It would be possible
to study other D-branes in these theories without too much trouble,
since the orbifold action has been fixed, for example one could study
the unstable branes which correspond to $S\in O(8)$ with $\det(S)=-1$.

\acknowledgments
We thank Andreas Recknagel for comments on the manuscript.  M.K. would
like to thank the ITP, Santa Barbara and the theory group at Harvard
University for hospitality.

\bibliographystyle{JHEP}
\bibliography{spires}

\providecommand{\href}[2]{#2}\begingroup\raggedright\begin{thebibliography}{10}

\bibitem{Vafa:1986wx}
C.~Vafa, {\it Modular invariance and discrete torsion on orbifolds},  {\em
  Nucl. Phys.} {\bf B273} (1986) 592.

\bibitem{Vafa:1995rv}
C.~Vafa and E.~Witten, {\it On orbifolds with discrete torsion},  {\em J. Geom.
  Phys.} {\bf 15} (1995) 189--214,
  [\href{http://xxx.lanl.gov/abs/hep-th/9409188}{{\tt hep-th/9409188}}].

\bibitem{Callan:1988wz}
J.~Callan, Curtis~G., C.~Lovelace, C.~R. Nappi, and S.~A. Yost, {\it Loop
  corrections to superstring equations of motion},  {\em Nucl. Phys.} {\bf
  B308} (1988) 221.

\bibitem{Polchinski:1988tu}
J.~Polchinski and Y.~Cai, {\it Consistency of open superstring theories},  {\em
  Nucl. Phys.} {\bf B296} (1988) 91.

\bibitem{DiVecchia:1999rh}
P.~Di~Vecchia and A.~Liccardo, {\it D branes in string theory. {I}},
  \href{http://xxx.lanl.gov/abs/hep-th/9912161}{{\tt hep-th/9912161}}.

\bibitem{DiVecchia:1999fx}
P.~Di~Vecchia and A.~Liccardo, {\it D-branes in string theory. {II}},
  \href{http://xxx.lanl.gov/abs/hep-th/9912275}{{\tt hep-th/9912275}}.

\bibitem{Recknagel:1998sb}
A.~Recknagel and V.~Schomerus, {\it D-branes in {Gepner} models},  {\em Nucl.
  Phys.} {\bf B531} (1998) 185--225,
  [\href{http://xxx.lanl.gov/abs/hep-th/9712186}{{\tt hep-th/9712186}}].

\bibitem{Douglas:1998xa}
M.~R. Douglas, {\it D-branes and discrete torsion},
  \href{http://xxx.lanl.gov/abs/hep-th/9807235}{{\tt hep-th/9807235}}.

\bibitem{Douglas:1999hq}
M.~R. Douglas and B.~Fiol, {\it D-branes and discrete torsion. ii},
  \href{http://xxx.lanl.gov/abs/hep-th/9903031}{{\tt hep-th/9903031}}.

\bibitem{Mukhopadhyay:1999pu}
S.~Mukhopadhyay and K.~Ray, {\it D-branes on fourfolds with discrete torsion},
  {\em Nucl. Phys.} {\bf B576} (2000) 152--176,
  [\href{http://xxx.lanl.gov/abs/hep-th/9909107}{{\tt hep-th/9909107}}].

\bibitem{Berenstein:2000hy}
D.~Berenstein and R.~G. Leigh, {\it Discrete torsion, {AdS/CFT} and duality},
  {\em JHEP} {\bf 01} (2000) 038,
  [\href{http://xxx.lanl.gov/abs/hep-th/0001055}{{\tt hep-th/0001055}}].

\bibitem{Gomis:2000ej}
J.~Gomis, {\it D-branes on orbifolds with discrete torsion and topological
  obstruction},  {\em JHEP} {\bf 05} (2000) 006,
  [\href{http://xxx.lanl.gov/abs/hep-th/0001200}{{\tt hep-th/0001200}}].

\bibitem{Klein:2000tf}
M.~Klein and R.~Rabadan, {\it Orientifolds with discrete torsion},  {\em JHEP}
  {\bf 07} (2000) 040, [\href{http://xxx.lanl.gov/abs/hep-th/0002103}{{\tt
  hep-th/0002103}}].

\bibitem{Aspinwall:2000xs}
P.~S. Aspinwall and M.~R. Plesser, {\it D-branes, discrete torsion and the
  {McKay} correspondence},  {\em JHEP} {\bf 02} (2001) 009,
  [\href{http://xxx.lanl.gov/abs/hep-th/0009042}{{\tt hep-th/0009042}}].

\bibitem{Aspinwall:2000xv}
P.~S. Aspinwall, {\it A note on the equivalence of {Vafa's} and {Douglas's}
  picture of discrete torsion},  {\em JHEP} {\bf 12} (2000) 029,
  [\href{http://xxx.lanl.gov/abs/hep-th/0009045}{{\tt hep-th/0009045}}].

\bibitem{Feng:2000mw}
B.~Feng, A.~Hanany, Y.-H. He, and N.~Prezas, {\it Discrete torsion, covering
  groups and quiver diagrams},  {\em JHEP} {\bf 04} (2001) 037,
  [\href{http://xxx.lanl.gov/abs/hep-th/0011192}{{\tt hep-th/0011192}}].

\bibitem{Sharpe:2000ki}
E.~R. Sharpe, {\it Discrete torsion},
  \href{http://xxx.lanl.gov/abs/hep-th/0008154}{{\tt hep-th/0008154}}.

\bibitem{Gutperle:2000bf}
M.~Gutperle, {\it Non-{BPS} {D-branes} and enhanced symmetry in an asymmetric
  orbifold},  {\em JHEP} {\bf 08} (2000) 036,
  [\href{http://xxx.lanl.gov/abs/hep-th/0007126}{{\tt hep-th/0007126}}].

\bibitem{Brunner:1999fj}
I.~Brunner, A.~Rajaraman, and M.~Rozali, {\it D-branes on asymmetric
  orbifolds},  {\em Nucl. Phys.} {\bf B558} (1999) 205--215,
  [\href{http://xxx.lanl.gov/abs/hep-th/9905024}{{\tt hep-th/9905024}}].

\bibitem{Diaconescu:1999dt}
D.-E. Diaconescu and J.~Gomis, {\it Fractional branes and boundary states in
  orbifold theories},  {\em JHEP} {\bf 10} (2000) 001,
  [\href{http://xxx.lanl.gov/abs/hep-th/9906242}{{\tt hep-th/9906242}}].

\bibitem{Diaconescu:2000ec}
D.-E. Diaconescu and M.~R. Douglas, {\it D-branes on stringy {Calabi-Yau}
  manifolds},  \href{http://xxx.lanl.gov/abs/hep-th/0006224}{{\tt
  hep-th/0006224}}.

\bibitem{Billo:2000yb}
M.~Billo, B.~Craps, and F.~Roose, {\it Orbifold boundary states from {Cardy's}
  condition},  {\em JHEP} {\bf 01} (2001) 038,
  [\href{http://xxx.lanl.gov/abs/hep-th/0011060}{{\tt hep-th/0011060}}].

\bibitem{Gaberdiel:1999ch}
M.~R. Gaberdiel and J.~Stefanski, Bogdan, {\it Dirichlet branes on orbifolds},
  {\em Nucl. Phys.} {\bf B578} (2000) 58--84,
  [\href{http://xxx.lanl.gov/abs/hep-th/9910109}{{\tt hep-th/9910109}}].

\bibitem{Gaberdiel:2000fe}
M.~R. Gaberdiel, {\it Discrete torsion orbifolds and {D-branes}},  {\em JHEP}
  {\bf 11} (2000) 026, [\href{http://xxx.lanl.gov/abs/hep-th/0008230}{{\tt
  hep-th/0008230}}].

\bibitem{Craps:2001xw}
B.~Craps and M.~R. Gaberdiel, {\it Discrete torsion orbifolds and {D-branes}.
  {II}},  {\em JHEP} {\bf 04} (2001) 013,
  [\href{http://xxx.lanl.gov/abs/hep-th/0101143}{{\tt hep-th/0101143}}].

\bibitem{Douglas:1996sw}
M.~R. Douglas and G.~Moore, {\it D-branes, quivers, and {ALE} instantons},
  \href{http://xxx.lanl.gov/abs/hep-th/9603167}{{\tt hep-th/9603167}}.

\bibitem{Bergman:2000ni}
O.~Bergman and M.~R. Gaberdiel, {\it On the consistency of orbifolds},  {\em
  Phys. Lett.} {\bf B481} (2000) 379--385,
  [\href{http://xxx.lanl.gov/abs/hep-th/0001130}{{\tt hep-th/0001130}}].

\bibitem{Cardy:1989ir}
J.~L. Cardy, {\it Boundary conditions, fusion rules and the {Verlinde}
  formula},  {\em Nucl. Phys.} {\bf B324} (1989) 581.

\end{thebibliography}\endgroup

\end{document}